\documentclass[11pt]{article}

\usepackage[english]{babel}
\usepackage[utf8]{inputenc}
\usepackage[T1]{fontenc}
\usepackage{lmodern}
\usepackage[margin=1in]{geometry}

\usepackage{amsmath}
\usepackage{amssymb}
\usepackage{bm}

\usepackage{graphicx}
\usepackage{booktabs}
\usepackage{tabularx}
\usepackage{longtable}
\usepackage{array}
\usepackage{float}
\usepackage{threeparttable}
\usepackage[section]{placeins}

\usepackage{ragged2e}
\usepackage{natbib}
\usepackage[hidelinks]{hyperref}

\renewcommand{\arraystretch}{0.95}

\newcommand{\paperabstract}{Online learning has increased the need to understand how student engagement patterns shape learning outcomes in flexible, technology-mediated environments. We propose a Bayesian nonparametric dynamic item response theory (IRT) framework for tracking within-individual ability trajectories across instructional units. The proposed model integrates B-spline basis expansions to capture nonlinear effects of engagement behaviors on ability drift, alongside a Mixture-of-Finite-Mixtures (MFM) prior to automatically determine the number of latent learner clusters. The framework addresses three limitations in the existing literature: (1) rigid linearity assumptions in engagement--ability relationships, (2) dependence on prespecified cluster counts, and (3) limited capacity to track longitudinal ability dynamics. We apply the model to longitudinal data from 198 undergraduates completing a 9-chapter introductory statistics course on CourseKata. The model automatically identified four distinct learner profiles: struggling-declining (11\%), low-stable (23\%), mainstream-stable (55\%), and high-improving (12\%). Results indicate that ability trajectories remained remarkably stable across chapters, and engagement quantity metrics did not significantly predict ability drift. These findings suggest that, in introductory online statistics education, academic ability primarily reflects a stable preexisting characteristic rather than a dynamically malleable course outcome. Ultimately, this framework offers a flexible tool for learner profiling to inform adaptive instructional design.

\vspace{1em}
\noindent\textbf{Keywords:} B-splines; educational assessment; engagement behaviors; mixture of finite mixtures
}

\begin{document}
\justifying

\thispagestyle{plain}
\begin{center}
    {\Large \bfseries
    Modeling Nonlinear Ability Trajectories and Learner Heterogeneity in Online Learning: A Bayesian Nonparametric Dynamic IRT Framework
    \par}
    \vspace{1em}
    Zhihua Ma\textsuperscript{1},
Alice Xu\textsuperscript{2},
Icy Zhang\textsuperscript{3},
Guanyu Hu\textsuperscript{4}

\vspace{1em}

\textsuperscript{1}Department of Statistics, Shenzhen University\\
\textsuperscript{2}Department of Psychology, University of California, Los Angeles\\
\textsuperscript{3}Department of Educational Psychology, University of Wisconsin-Madison\\
\textsuperscript{4}Department of Statistics \& Probability, Michigan State University
\end{center}


\begin{abstract}
\noindent
\paperabstract
\end{abstract}


\section{Introduction}
\label{sec:intro}
Online learning platforms, digital textbooks, and
learning management systems have increasingly shifted learning into
flexible and technology-mediated environments
\citep{means2014learning,siemens2013learning,bonk2009world}.
These platforms expand access and support more personalized learning
experiences, but they also expose substantial differences in students'
self-regulation, motivation, persistence, and awareness of their own
progress \citep{winters2011motivation,broadbent2015self}.
Consequently, understanding learning in online settings requires more
than evaluating final outcomes; it requires examining how students
engage with digital materials and how their knowledge develops over
time.

A key feature of online learning environments is their ability to
generate detailed process data that were largely unavailable in
traditional educational settings. Beyond quiz and test responses,
digital platforms can record when students access content, how long
they spend on particular materials, whether they revisit earlier
chapters, and how their response patterns change across repeated
learning opportunities
\citep{greller2012learning,romero2013data,baker2016educational}.
These fine-grained behavioral traces make it possible
to study learning as a dynamic process and are especially useful for
research on self-regulated learning, engagement, persistence, and
individual differences
\citep{zimmerman2002becoming,panadero2017review}.
By combining behavioral engagement with performance data, researchers
can obtain a richer picture of how learning unfolds, when students
struggle, and how educational systems might better support progress.

At the same time, these new forms of data introduce important
statistical challenges. Online learning data are often noisy,
irregularly spaced, highly individualized, and heterogeneous across
learners. Methods developed for cross-sectional educational testing are
not well suited to such settings because they typically focus on
end-of-test performance and treat ability as fixed during the assessment
period. To fully leverage online learning data, statistical methods must
accommodate temporal change, between-student heterogeneity, and the
complex behavioral information generated by digital platforms
\citep{vanlehn2005andes,baker2009state}.

Traditional item response theory (IRT) provides the foundation for
psychometric modeling by linking observed item responses to latent
learner ability and item characteristics
\citep{Lord1980,embretson2013item,rasch1960probabilistic}.
Although IRT offers interpretable measures of proficiency and supports
principled assessment design, standard formulations assume that ability
remains constant during the testing period. This assumption is often
reasonable in one-time examinations, but it becomes restrictive in
online learning environments, where learning is continuous and ability may
change across practice opportunities, instructional exposures, and
periods of inactivity. In such contexts, a static view of ability is
insufficient for modeling learning as it unfolds.

To address this limitation, dynamic item response (DIR) models extend
IRT by allowing latent ability to vary over time. These models typically
adopt a state-space formulation in which an observation equation links
current ability to item responses and a state equation describes how
ability evolves across learning occasions
\citep{martin2002dynamic,wang2013bayesian}. This framework is
especially appealing for online education because it treats learning as
an evolving latent process rather than a fixed trait.
Bayesian approaches further strengthen this framework by providing a
coherent way to estimate latent states, quantify uncertainty, and
accommodate irregular observation schedules and hierarchical dependence
structures \citep{fox2010bayesian,wang2013bayesian}.

Despite these advances, important methodological gaps remain. Existing
dynamic psychometric models often rely on restrictive parametric
assumptions about how learning changes over time, even though
trajectories in online environments may involve abrupt gains, plateaus,
regressions, or substantial irregularity. In addition, many models do
not explicitly account for latent heterogeneity in the learner
population, despite clear evidence that students may follow
qualitatively different learning pathways. Finally, online learning
platforms generate rich behavioral information, and statistical models
should be able to incorporate such predictors flexibly rather than
through overly simplified linear forms
\citep{rost1990rasch,richardson1997bayesian,ruppert2003semiparametric,
eilers1996flexible}. For example, additional time spent on the digital
learning platform may be beneficial up to a point; however, excessive
time may also indicate confusion with the content.
These limitations motivate Bayesian methods that can
jointly represent dynamic ability change, latent learner heterogeneity,
and nonlinear engagement effects.

This paper develops a Bayesian framework for online learning data that
integrates psychometric measurement with dynamic longitudinal modeling,
latent heterogeneity, and flexible covariate effects. Methodologically,
our contribution is to extend dynamic item response modeling in ways
that are better suited to online learning settings, where trajectories are
irregular, learner populations are heterogeneous, and the effects of
behavioral predictors may be nonlinear. Substantively, our work
contributes to applied educational psychology by connecting latent
learning progression with behavioral engagement in digital textbooks,
thereby providing a more nuanced understanding of how students learn in
online environments. 

The remainder of this article is organized as follows.
The ``Background and Related Literature'' section
reviews dynamic IRT models, mixture approaches, and spline-based psychometric
methods. The ``Proposed Model'' section develops the proposed Bayesian nonparametric dynamic IRT model. The ``Posterior Inference'' section
presents posterior inference. The ``Simulation Study'' section reports simulation studies evaluating empirical performance, the ``Empirical Application'' section applies the proposed method to longitudinal data from an online statistics course, and the ``Discussion'' section
concludes.


\section{Background and Related Literature}
\label{sec:background}

Item response theory (IRT) has long provided a principled framework for
modeling the relationship between latent proficiency and observed item
responses. In its classical form, IRT assumes that each learner has a
relatively stable latent ability and that the probability of a correct
response depends on that ability together with item characteristics such
as difficulty
\citep{rasch1960probabilistic,hambleton1991fundamentals,
embretson2013item}. A central strength of this framework is that it
estimates learner proficiency and item characteristics separately:
learners can be characterized in terms of their underlying ability,
while items can be characterized in terms of how difficult or
discriminating they are.
This separation of person and item parameters has made IRT foundational
in educational measurement, as it supports score comparability across
learners and assessments and provides an interpretable basis for
assessment design. These assumptions are especially appropriate in
traditional testing contexts, where responses are collected within a
short time window and learners' proficiency can reasonably be treated as
fixed during the assessment.

However, these assumptions become less tenable in longitudinal, online
learning environments. Online platforms make it possible to observe
students' learning processes repeatedly over time, including both their
item responses and their patterns of engagement with instructional
materials.  In such contexts, learner
proficiency is not static; it evolves as students practice, forget,
revisit material, receive feedback, engage with different resources, and
respond to interventions. Moreover, although online platforms generate
rich behavioral and response data, these data are often sparse,
unbalanced across learners and items, and locally dependent within
learning sessions. These features call for dynamic psychometric models
that move beyond treating proficiency as fixed and instead explicitly
model how learner proficiency changes over time.

Online learning environments are also characterized by
changing patterns of engagement. Because students often regulate their
own motivation, metacognition, and study behavior, engagement traces may
reflect processes related to self-regulated learning
\citep{pintrich2000multiple,zimmerman2002becoming,prasse2024challenges}.
From this perspective, engagement traces are not merely behavioral logs;
when theoretically grounded, they can serve as time-varying predictors
of ability drift within a dynamic IRT model \citep{mair2022bayesian}.
At the same time, learning analytics research cautions that trace data
should not be treated as direct measures of SRL without theoretical
grounding \citep{winne_learning_2017}. Accordingly, engagement
behaviors should be linked to an explicit learning model so that they
can be interpreted as meaningful indicators of how students regulate
learning and how those regulatory processes shape proficiency
trajectories.

Dynamic IRT models address this issue by allowing latent ability to vary
across time points \citep{martin2002dynamic,wang2013bayesian,
sun2026bayesian}. Instead of treating proficiency as fixed, these models
embed ability within a stochastic process so that repeated item
responses are linked to an evolving latent trait. This perspective is
particularly relevant in online learning contexts, where the goal is not
only to measure current performance but also to understand how learning
develops as students engage with the textbook or instructional content.

Bayesian methods have played a particularly important role in the
development of dynamic IRT. Because these models are hierarchical and
involve multiple sources of uncertainty, including latent abilities,
item parameters, subject-specific effects, and temporal innovations,
Bayesian inference offers a coherent way to estimate all model
components jointly \citep{fox2010bayesian,wang2013bayesian}. It also
facilitates the incorporation of prior information and partial pooling,
which is valuable when some students have dense response histories while
others contribute only sparse observations. 

An influential example is the Bayesian dynamic item response (DIR) model
of \citet{wang2013bayesian}, which extends IRT to account for irregular
time intervals, local dependence, and stochastic changes in ability over
time. Compared with static IRT, this framework is better aligned with
repeated-response educational data because it recognizes both the
temporal evolution of proficiency and the clustered nature of responses
collected within the same day or assessment context.

Nevertheless, existing Bayesian dynamic IRT models
\citep{wang2013bayesian} still face three limitations when applied to
online learning environments \citep{winne_learning_2017}. First, many
models impose relatively rigid parametric assumptions on how ability
evolves over time. In practice, learning trajectories in online settings
are often complex and uneven. Students may show abrupt improvement after
feedback, temporary regression after periods of inactivity, extended
plateaus, or highly individualized patterns of progress
\citep{Ramsay1991}. 
Simple parametric growth structures may therefore be insufficient for
capturing the nonlinear, fluctuating, and individualized nature of
learning in online environments \citep{kew2022learning}.

Second, standard dynamic IRT models generally do not explicitly
represent latent subgroups of learners. Although they may allow
continuous variation in initial proficiency or growth parameters, they
often assume that all learners follow variations of the same basic
developmental process. In online learning, this assumption may be
unrealistic. Learners can differ not only in degree but also in kind:
some may be consistently engaged, some may improve only after repeated
exposure, and others may exhibit unstable progress because of irregular
participation. Identifying such latent subpopulations is important for
scalable personalization, which motivates mixture-based psychometric
approaches such as mixture Rasch models and latent class IRT models
\citep{rasch1960probabilistic,mccutcheon1987latent,bacci2014class}.
Bayesian clustering methods provide a natural extension for modeling
this heterogeneity.
Finite mixture models are interpretable but require the
number of clusters to be specified in advance
\citep{richardson1997bayesian,fruhwirth2006finite}.
In particular, the Dirichlet process mixture model (DPMM) allows the
number of occupied clusters to grow with the data by placing a
Dirichlet process prior on the mixing distribution
\citep{ferguson1973bayesian,antoniak1974mixtures,escobar1995bayesian}.
However, DPMMs may favor partitions with many small clusters, which can
be undesirable when the goal is to identify a moderate number of
interpretable learner types. The mixture of finite mixtures (MFM)
framework offers an alternative by placing an explicit prior on the
unknown number of mixture components, thereby combining the
interpretability of finite mixtures with the adaptability of Bayesian
random-partition models
\citep{miller2018mixture,betancourt2022random,hu2024nonparametric,
pan2024bayesian}.

A third challenge is that the effects of predictors in online learning are
often nonlinear. Variables such as cumulative practice, time between
sessions, review frequency, intervention intensity, or prior exposure
may influence learning through threshold effects, diminishing returns,
or saturation. Linear and low-dimensional parametric specifications may
therefore fail to capture these relationships adequately. This has
motivated the use of flexible semiparametric regression tools,
particularly spline-based methods, which allow the shape of a covariate
effect to be learned from the data rather than imposed in advance
\citep{ruppert2003semiparametric,deBoor1978,eilers1996flexible}.
Among these tools, B-splines are especially attractive because of their
local support, numerical stability, and computational efficiency,
making them well suited for Bayesian hierarchical models of learning.

Taken together, this literature underscores the need for more flexible
Bayesian methods that can capture the longitudinal and complex nature of
knowledge development, particularly in online, self-regulated learning
environments. Modern online learning data require approaches capable of
accommodating heterogeneous learner populations, irregular and nonlinear
developmental trajectories, and rich behavioral predictors.
The proposed framework integrates these components
within a Bayesian dynamic IRT model by combining latent learner
clustering with spline-based engagement effects on ability drift.

\section{Proposed Model}
\label{sec:model}

In this section, a Bayesian nonparametric dynamic IRT model is developed to circumvent the limitations identified above 
To be specific, we employ a dynamic IRT process to track within-individual ability trajectories across instructional chapters, a B-spline basis expansions to capture nonlinear engagement-ability drift, and an MFM prior to infer the number of latent learner clusters jointly with cluster-specific parameters. All these three components are integrated within a hierarchical Bayesian framework. 

Let $i = 1,\ldots,N$ denote the $i$th student,
$t = 1,\ldots,T$ denote the $t$th instructional chapter,
$j = 1,\ldots,J_t$ denote the $j$th items within chapter $t$,
and $k = 1,\ldots,K$ denote the $k$th latent learner cluster.
The observed data consist of binary item responses
$Y_{ijt} \in \{0,1\}$ and $P$-dimensional chapter-level engagement
covariates per student $x_{it}^{(p)} (p=1,\cdots,P)$.

\subsection{Observation Model}
\label{subsec:obs_model}

Conditional on latent ability $\theta_{it}$ and item difficulty
$b_{jt}$, binary responses are modeled via the one-parameter
logistic (1PL) IRT model \citep{rasch1960probabilistic}:
\begin{equation}
  Y_{ijt} \mid \theta_{it},\, b_{jt}
  \;\overset{\mathrm{ind}}{\sim}\;
  \mathrm{Bernoulli}\bigl(p_{ijt}\bigr),\qquad
  \mathrm{logit}(p_{ijt}) = \theta_{it} - b_{jt}.\label{eq:obs}
\end{equation}
Conditional independence of responses given the latent parameters is assumed throughout
\citep{embretson2013item}.
The 1PL formulation constrains item discrimination to unity,
which we adopt to balance model parsimony with the
available sample size ($N = 198$).

\paragraph{Item difficulty.}
Raw item difficulties are drawn from a chapter-specific
normal distribution:
\begin{equation}
  b_{jt}^{\mathrm{raw}} \;\sim\; \mathcal{N}(0,\;\sigma_{b,t}^{2}),
  \qquad j = 1,\ldots,J_t,
  \label{eq:b_raw}
\end{equation}
and are centered within each chapter to resolve additive
non-identifiability:
\begin{equation}
  b_{jt} = b_{jt}^{\mathrm{raw}} - \bar{b}_{t},
  \qquad
  \bar{b}_{t} = \frac{1}{J_t}
  \sum_{j=1}^{J_t} b_{jt}^{\mathrm{raw}}.
  \label{eq:b_center}
\end{equation}

A weakly informative half-normal prior is assumed for the chapter-level difficulty standard deviation $\sigma_{b,t}$:
\begin{equation}
  \sigma_{b,t} \;\sim\; \mathcal{N}^{+}(0,\;1),
  \qquad t = 1,\ldots,T.
  \label{eq:prior_sdb}
\end{equation}
This prior assigns negligible probability to implausibly large
difficulty ranges ($\sigma_{b,t} > 4$) while placing no
meaningful constraint on the range observed in practice.

\subsection{Ability Dynamics}
\label{subsec:ability_dynamics}

Student abilities are modeled as latent states evolving
according to a first-order Markov process across chapters
\citep{Durbin2012}.
At the initial chapter, ability is anchored to the
cluster-specific mean:
\begin{equation}
  \theta_{i1} = \theta_{m,k} + \varepsilon_{i1},
  \qquad z_i = k,
  \qquad
  \varepsilon_{i1} \sim \mathcal{N}\!\left(0,\;\tau_\theta^{-1}\right),
  \label{eq:theta_init}
\end{equation}
where $z_i \in \{1,\ldots,K\}$ denotes the latent cluster
membership of student $i$, $\theta_{m,k}$ is the mean ability
of cluster $k$, and $\tau_\theta$ is a shared precision
parameter that governs longitudinal stability.
For $t = 2,\ldots,T$, ability transitions follow
\begin{equation}
  \theta_{it} = \theta_{i,t-1} + \delta_{it} + \varepsilon_{it},
  \qquad
  \varepsilon_{it}
  \overset{\mathrm{iid}}{\sim}
  \mathcal{N}\!\left(0,\;\tau_\theta^{-1}\right),
  \label{eq:theta_trans}
\end{equation}
where $\delta_{it}$ denotes the \textit{expected ability drift}
driven by engagement behavior,
and $\varepsilon_{it}$ captures unsystematic within-individual
fluctuations.
The precision $\tau_\theta$ is shared across all students and
chapters, encoding the assumption that trajectories exhibit a
homogeneous degree of temporal smoothness;
equivalently, the corresponding process standard deviation
$\sigma_\theta = \tau_\theta^{-1/2}$ represents the
chapter-to-chapter stochastic variation in ability beyond
systematic drift. A normal prior and a Gamma prior is placed on $\theta_{m,k}$ and $\tau_\theta$, respectively.


\paragraph{Nonlinear Engagement-Ability Drift via B-Splines.}
The linearity assumption that pervades existing dynamic IRT models is rarely theoretically motivated.
We relax this assumption by specifying the engagement-drift
relationship through B-spline basis expansions
\citep{deBoor1978, eilers1996flexible}, allowing the model to learn the
functional form from the data.

For the $p$th engagement covariate, let
$\mathbf{B}^{(p)}(x) =
\bigl(B_{p1}(x),\ldots,B_{pL}(x)\bigr)^{\top}$
denote a B-spline basis of degree 3 with $L = 3$ basis functions.
Interior knots are placed at the empirical tertiles of
$x_{it}^{(p)}$ pooled across all students and chapters,
ensuring that the basis spans the observed covariate
support with roughly equal density.
Each basis function is normalized to have unit $L_2$-norm
over the empirical support prior to estimation, which
standardizes the scale of the associated spline
coefficients. 

The smooth engagement-drift function for the $p$th covariate is
\begin{equation}
  f_p(x) = \sum_{l=1}^{L} b_{pl}\,B_{pl}(x)= \mathbf{b}_p^{\top}\mathbf{B}^{(p)}(x),
  \label{eq:spline_func}
\end{equation}
where $\mathbf{b}_p = (b_{p1},\ldots,b_{pL})^{\top}$
are global spline coefficients shared across students and
clusters.
The functions $f_p(\cdot)$ thus capture the
shape of the engagement-ability relationship at the population
level.

\paragraph{Cluster-specific drift.}
The expected ability drift for student $i$ in cluster $k$
at chapter $t$ is
\begin{equation}
  \delta_{it}
  = \sum_{p=1}^{P} \beta_{p,k}\,f_p\!\left(x_{it}^{(p)}\right),
  \qquad z_i = k,
  \label{eq:drift}
\end{equation}
where $\beta_{p,k} \in \mathbb{R}$ is a cluster-specific
scaling coefficient.
This two-level specification separates the
\textit{shape} of the engagement function
(global, estimated by $\mathbf{b}_p$) from its
\textit{magnitude and sign} within each cluster
(estimated by $\beta_{p,k}$), enabling the model to
represent qualitatively heterogeneous engagement responses
across learner profiles.
When $\beta_{p,k} = 0$ for all $k$, the model reduces to a
dynamic IRT model without covariate-driven drift. 

Weakly informative normal priors are assigned to the
global spline coefficients and cluster scaling factors:
\begin{align}
  b_{pl}&\;\overset{\mathrm{iid}}{\sim}\;
    \mathcal{N}(0,\;0.5^{2}),
  \quad l = 1,\ldots,L,\;\;
  \label{eq:prior_bspline} \\[4pt]
  \beta_{p,k}
    &\;\overset{\mathrm{iid}}{\sim}\;
    \mathcal{N}(0,\;2^{2}),
  \quad k = 1,\ldots,K\;\;.
  \label{eq:prior_beta}
\end{align}
The tighter prior on $\mathbf{b}_p$
provides mild regularization against implausible
oscillations in the spline shape, consistent with the
P-spline literature \citep{eilers1996flexible, Lang2004}.
The broader prior on $\beta_{p,k}$
imposes minimal constraint on the direction or magnitude
of engagement effects at the cluster level.

\subsection{Mixture-of-Finite-Mixtures Prior for Automatic Cluster Enumeration}
\label{subsec:mfm}

To address the limitation of prespecified $K$, we adopt the MFM prior \citep{miller2018mixture},
which treats $K$ as a random variable and enables joint
Bayesian inference on the number and composition of latent
learner clusters.

Let $M$ denote an upper bound on the number of components.
The MFM is defined as
\begin{align}
  K &\;\sim\;p_K(k)\;\propto\;
    \mathrm{Poisson}(k;\,\lambda=1)\,\mathbf{1}(1\leq k\leq M),
  \label{eq:mfm_K} \\[4pt]
  (\pi_1,\ldots,\pi_K) \mid K&\;\sim\; \mathrm{Dirichlet}(\gamma,\ldots,\gamma),
  \label{eq:mfm_pi} \\[4pt]
  z_i \mid \boldsymbol{\pi}
    &\;\overset{\mathrm{iid}}{\sim}\;
    \mathrm{Categorical}(\boldsymbol{\pi}),
  \quad i=1,\ldots,N,
  \label{eq:mfm_z}
\end{align}
where $\gamma =1$ is the symmetric Dirichlet concentration
parameter.
The Poisson$(1)$ component prior places a priori preference
on parsimonious solutions while not excluding larger $K$
values.
In practice, the MFM posterior concentrates tightly around
the data-supported number of active clusters; components
assigned negligible mixing weight ($\pi_k \approx 0$) are
effectively pruned \citep{miller2018mixture,Fruhwirth2019}.
Each active cluster $k$ is characterized by a mean ability
$\theta_{m,k}$ and engagement scaling coefficients
$\beta_{p,k}$.

The MFM prior offers two key advantages over the Dirichlet Process
Mixture \citep{escobar1995bayesian}. First, posterior inference on $K$ is consistent: as $N\to\infty$, the posterior probability assigned to the true
$K^*$ converges to one \citep{miller2018mixture}, whereas DPM
posteriors over-segment data because the expected number
of occupied clusters grows as $O(\log N)$ \citep{Miller2014}.
Second, the MFM framework produces interpretable and reproducible cluster solutions without requiring sequential model comparisons across candidate $K$ values, thereby avoiding the information-criterion sensitivity and
computational redundancy.

\subsection{Identifiability}
\label{subsec:identifiability}

Two sources of non-identifiability require explicit treatment.

\paragraph{Additive translation invariance.}
The likelihood in Equation~\eqref{eq:obs} is invariant under
the transformation
$\theta_{it} \mapsto \theta_{it} + c$,
$b_{jt} \mapsto b_{jt} + c$ for any constant
$c \in \mathbb{R}$.
We resolve this by imposing the within-chapter centering
constraint of Equation~\eqref{eq:b_center}, which fixes
$\bar{b}_t = 0$ for each chapter.
Under this normalization, the origin of the ability scale is
defined by the average item difficulty within each chapter,
and chapter-level differences in absolute difficulty are
absorbed into the cluster means $\theta_{m,k}$.
The uncentred chapter-mean difficulties
$\bar{b}_t^{\mathrm{raw}}$ serve as indices of chapter-level
difficulty in subsequent analyses.

\paragraph{Label switching.}
Mixture models are invariant to permutations of cluster
labels, rendering the posterior of cluster-specific
parameters multimodal and uninterpretable without
post-processing \citep{Stephens2000}.
We apply the \textit{Equivalence Classes Representatives}
(ECR) relabeling algorithm \citep{Papastamoulis2016}
to the raw MCMC output.
ECR identifies, for each posterior draw, the label
permutation that minimises within-chain variation with
respect to a reference iteration, yielding a relabeled
chain that concentrates mass on a single mode.
After relabeling, clusters are sorted in ascending order
of posterior mean ability $\hat{\theta}_{m,k}$, so that
$k=1$ consistently denotes the lowest-ability profile
and $k=K$ the highest throughout the paper.

\section{Posterior Inference}
\label{sec:inference}

For the proposed model, closed-form posterior distributions are not available.
We therefore resort to Markov chain Monte Carlo (MCMC) simulation,
implemented in the \texttt{NIMBLE} probabilistic programming
framework \citep{deValpine2017} for \textsf{R}\citep{RCoreTeam2024}.
\texttt{NIMBLE} compiles model code to \texttt{C++} at runtime,
providing near-native execution speed while retaining the
flexibility of a declarative model specification.

\subsection{MCMC Sampling Strategy}
\label{subsec:mcmc_strategy}

A key challenge in MCMC for the proposed model is the high
posterior correlation among the latent ability sequences
$\{\theta_{it}\}_{t=1}^{T}$, the cluster means
$\boldsymbol{\theta}_m$, and the item difficulty parameters
$\{b_{jt}\}$.
Naive element-wise Metropolis-Hastings (MH) sampling for
these blocks yields poor mixing due to strong posterior
dependencies \citep{Roberts1997}.
We address this by tailoring the sampler type to the structural
role of each parameter block.

\paragraph{Latent ability trajectories $\{\theta_{it}\}$.}
For each student $i$, the joint trajectory
$(\theta_{i1},\ldots,\theta_{iT})^{\top}$ is sampled using
a \textit{slice sampler} \citep{Neal2003} applied
independently to each element $\theta_{it}$.
Slice sampling is particularly well-suited here because the
full conditional of $\theta_{it}$ is log-concave, and the
slice sampler is guaranteed to mix well for log-concave targets
without requiring manual tuning of a proposal scale.
Each $\theta_{it}$ is updated conditionally on all other
ability states, current item difficulties, and the student's
cluster assignment, cycling over $i = 1,\ldots,N$ and
$t = 1,\ldots,T$ within each iteration.

\paragraph{Global B-spline coefficients
$\mathbf{b}_p$.}
The spline coefficients are sampled jointly using a
\textit{random-walk block Metropolis--Hastings} (RWB-MH)
sampler \citep{Roberts1997}.
Joint updating of the full coefficient vector is essential
because the B-spline basis functions
$B_{sl}(\cdot)$ are correlated by construction, inducing strong posterior dependence among the corresponding coefficients.

\paragraph{Cluster-specific scaling coefficients
$\{\beta_{p,k}\}$.}
For each cluster $k$ and engagement type $p$, the coefficient
$\beta_{p,k}$ is updated with an independent
\textit{random-walk MH} sampler.
Because $\beta_{p,k}$ enters the drift linearly, its full conditional is available in semi-closed form up to a normalizing constant involving
the logistic likelihood, making standard MH efficient in
practice.

\paragraph{Cluster membership indicators $\{z_i\}$.}
Each cluster assignment $z_i$ is updated via
\textit{collapsed Gibbs sampling} \citep{Liu1994}.
Specifically, the mixing weights $\boldsymbol{\pi}$ are
marginalized out analytically using the Dirichlet-Categorical
conjugacy, and $z_i$ is drawn from the resulting
discrete full conditional:
\begin{equation}
  p(z_i = k \mid \cdot\,) \;\propto\;
  \frac{n_{-i,k} + \gamma}{N -1 + K\gamma}\;
  \cdot\;
  p\!\left(\theta_{i1},\ldots,\theta_{iT}\mid z_i = k,\; \boldsymbol{\theta}_m,\; \tau_\theta,\;
    \boldsymbol{\beta},\; \mathbf{b}_p\right),\label{eq:collapsed_gibbs}
\end{equation}
where $n_{-i,k}$ is the number of students other than $i$
currently assigned to cluster $k$, and
$\gamma = 1$ is the Dirichlet concentration parameter.

\paragraph{Cluster means $\boldsymbol{\theta}_m$ and
hyperparameters.}
Cluster means $\theta_{m,k}$ and the process precision
$\tau_\theta$ are updated with independent random-walk MH
samplers, as are the chapter-level item difficulty standard deviation
$\sigma_{b,t}$.
The raw item difficulties $b_{jt}^{\mathrm{raw}}$ are updated
using slice samplers, exploiting the same log-concavity
argument as the ability parameters.

\paragraph{Number of components $K$.}
Under the MFM prior, the number of active components is
not updated directly; instead, it is computed at each
iteration as the number of clusters to which at least one
student is assigned:
\begin{equation}
  K_{\mathrm{est}}^{(r)} = \bigl|\{k : n_k^{(r)} \geq 1\}\bigr|,
  \quad r = 1,\ldots,R,
  \label{eq:K_est}
\end{equation}
where $n_k^{(r)} = \sum_{i=1}^{N}\mathbf{1}(z_i^{(r)} = k)$
is the cluster size at iteration $r$.
The posterior distribution of $K$ is then summarised from
the resulting sequence $\{K_{\mathrm{est}}^{(r)}\}_{r=1}^{R}$,
with the modal value taken as the point estimate
$\hat{K}$.

All analyses used 40,000 total MCMC iterations, with the first 10,000 discarded as burn-in and the remaining 30,000 thinned by a factor of 10, yielding $R = 3,000$ posterior draws per parameter. Convergence is guaranteed via traceplots. 

\subsection{Label Switching Correction via ECR}
\label{subsec:ecr}

Mixture posteriors are invariant under permutations of the
cluster labels, producing a multimodal MCMC output that
cannot be directly summarized by standard posterior means
or credible intervals.
We apply the \textit{Equivalence Classes Representatives}
(ECR) algorithm \citep{Papastamoulis2016} to the output after burn-in.

The ECR procedure operates as follows. A \textit{pivot} iteration $r^*$ is selected as the draw
whose cluster assignment vector $\mathbf{z}^{(r^*)}$ is
most representative of the modal partition, determined by
maximising the within-chain pairwise co-clustering
frequency.
For each subsequent draw $r$, the optimal label permutation
$\hat{\sigma}^{(r)}$ is found by solving the linear
assignment problem
\begin{equation}
  \hat{\sigma}^{(r)}
  = \operatorname*{arg\,min}_{\sigma \in \mathcal{S}_K}
  \sum_{i=1}^{N}\mathbf{1}
  \!\left(z_i^{(r,\,\sigma)} \neq z_i^{(r^*)}\right),\label{eq:ecr}
\end{equation}
where $\mathcal{S}_K$ denotes the set of all permutations
of $\{1,\ldots,K\}$ and
$z_i^{(r,\,\sigma)} = \sigma(z_i^{(r)})$.
The Hungarian algorithm \citep{Kuhn1955} solves
Equation~\eqref{eq:ecr} in $O(K^3)$ time, making the
procedure computationally negligible relative to MCMC
runtime.
Following relabeling, all cluster-specific parameters
are permuted accordingly, and clusters are finally sorted
in ascending order of posterior mean ability.

\section{Simulation Study}
\label{sec:simulation}

To evaluate the empirical performance of the proposed
model, we conducted a Monte Carlo simulation study
with 100 independent replications per experimental condition.
The study addressed three primary questions:
\begin{enumerate}
  \item Can the MFM prior reliably recover the true number of
        latent learner clusters ($K^* = 3$) without
        pre-specification?
  \item Is the nonlinear engagement--ability drift function
        recovered more accurately by the B-splines specification
        than by a linear alternative when the true relationship
        is genuinely nonlinear?
  \item Are individual ability parameters $\theta_{it}$ estimated
        with adequate precision under both specifications?
\end{enumerate}

\subsection{Data-Generating Process}
\label{subsec:sim_dgp}

Data were generated to mimic the structure of the empirical
application: $N = 198$ students, $T = 9$ instructional chapters,
and chapter-level item counts matching the CourseKata course
($J_t \in \{13,\,32,\,31,\,32,\,32,\,30,\,12,\,20,\,29\}$,
$\sum_t J_t = 230$).

\paragraph{Cluster structure.}
Three equally-sized latent clusters were specified
($K^* = 3$, $n_1 = n_2 = n_3 = 66$), with student
assignments $z_i \in \{1,2,3\}$ drawn uniformly.
Cluster-specific mean abilities were fixed at $\boldsymbol{\theta}_m^* = (-3,\;0,\;+3)^\top$,
spanning a range of 6 logit units and representing three
qualitatively distinct learner profiles: low-ability,
average-ability, and high-ability.
Initial abilities were drawn as $\theta_{i1} \sim \mathcal{N}\!\left(\theta_{m,z_i}^*,\;
  \sigma_\theta^{*2}\right)$
with the same process standard deviation $\sigma_\theta^* = 0.5\;\;(\tau_\theta^* = 4)$ governing all
chapter-to-chapter transitions.

\paragraph{Engagement covariates.}
Two chapter-level covariates were generated
independently for each student-chapter combination. We assume $x_{it}^{(1)} \sim \mathcal{N}(1,\;0.4^2), x_{it}^{(2)} \sim \mathcal{N}(0,\;0.6^2)$, and 
the mean of $x_{it}^{(1)}$ was set to $\mu = 1$ to ensure that the cubic function $f_1(x) = x^3$ takes on clearly non-zero and asymmetric values over the realistic support $[0.07,\,1.93]$, providing a stringent test of the spline's ability to detect curvature.
The standard deviation of $x_{it}^{(2)}$  was chosen so that $\exp(x)$ varies by a factor of up to 8 across the support $[-1.8,\,1.8]$.

\paragraph{True nonlinear drift functions.}
The true engagement-ability drift functions were specified as $f_1^*(x) = x^3$, and $f_2^*(x) = \exp(x)$, 
both of which are smooth but clearly non-linear: $f_1^*$
is a monotone cubic with pronounced curvature for
$|x| > 0.5$, and $f_2^*$ is convex with exponentially
increasing slope.

\paragraph{Cluster-specific drift.}
The expected ability drift was specified as
\begin{equation}
  \delta_{it}^* = \beta_{1,k}^*\,f_1^*\!\left(x_{it}^{(1)}\right)
  + \beta_{2,k}^*\,f_2^*\!\left(x_{it}^{(2)}\right),
  \qquad z_i = k,
  \label{eq:sim_drift}
\end{equation}
with cluster-specific scaling coefficients
\begin{equation}
  \beta_{1,k}^* = \beta_{2,k}^* =
  \begin{cases}
    +0.15 & k = 3 \;\text{(high-ability, positive drift)}\\
    \phantom{+}0 & k = 2\;\text{(average-ability, zero drift)}\\
    -0.15 & k = 1\;\text{(low-ability, negative drift)}
  \end{cases}.
  \label{eq:sim_beta}
\end{equation}
This design embeds a substantively important test:
cluster $k = 2$ has \emph{zero drift by construction}
($\delta_{it}^* \equiv 0$ for all $i \in k=2$), so the model
must correctly infer that the middle cluster exhibits no
engagement-ability relationship.
The scaling coefficients were calibrated to produce a
cumulative drift of approximately $\pm 3.2$ logit units over
$T-1 = 8$ chapters for the extreme clusters (G1 and G3),
matching the between-cluster range of 6 logit units.

\paragraph{Item parameters and response generation.}
Item difficulties were drawn as
$b_{jt}^{\mathrm{raw}} \sim \mathcal{N}(0,\,\sigma_b^2)$
with $\sigma_b = 0.5$ for all chapters, and centered within
each chapter.
Binary responses were generated from the 1PL model
(Equation~\ref{eq:obs}).
All 100 replications used the same structural parameters;
only the random draws of $z_i$, $\theta_{i1}$, $x_{it}^{(p)}$,
$b_{jt}^{\mathrm{raw}}$, and $\varepsilon_{it}$ varied
across replications.

\subsection{Evaluation Criteria}
\label{subsec:sim_criteria}

For clustering performance, we calculate the Adjusted Rand Index (ARI; \citealt{Hubert1985}), the $K$ recovery rate ($P(\hat{K} = K^*)$), Precision and Recall for evaluation.

For ability recovery, we calculate the Mean absolute bias (MAB) and coverage probability (CP\textsubscript{95}) for $\theta_{it}$. 
Since the cluster-specific scaling coefficients
$\beta_{p,k}$ and the global spline coefficients $\mathbf{b}_p$
are scale-unidentifiable, the \emph{drift} is adopted as the primary identifiable evaluation quantity, and we report its MAB and the correlation for the cluster-average drift trajectory. 

For item difficulty recovery, we also report the MAB and  CP\textsubscript{95} for $b_{jt}$.


\subsection{Performance of the Proposed Model}
\label{subsec:sim_proposed}

\subsubsection{Cluster Recovery}
\label{subsubsec:sim_cluster}

The cluster recovery indices are obtained through averaging across 100 replications. 
The MFM prior correctly identified $K^* = 3$ active clusters
in \textbf{90\%} of replications, with a mean ARI of $0.897$
($\mathrm{SD} = 0.109$), precision of $0.911$, and recall of
$0.906$.
These results confirm that automatic cluster enumeration is
reliable in samples of the present size without pre-specifying
the number of components.

\begin{figure}
\centering
\includegraphics[scale=0.6]{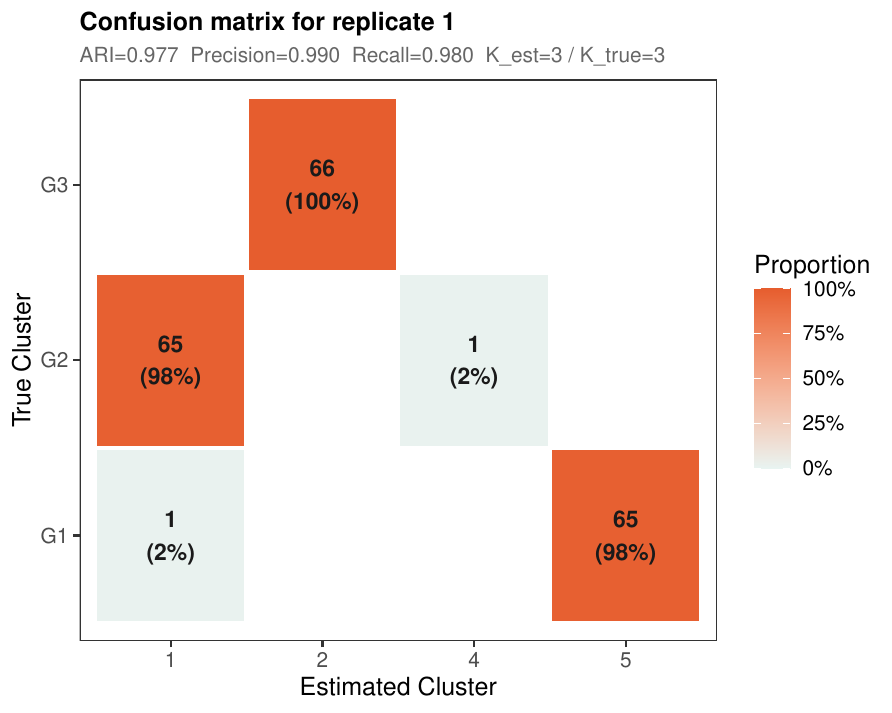}
\caption{\label{fig:sim_confusion}Cluster Assignment Confusion Matrix: Representative Single Replication of the Proposed Model.}
\end{figure}

Figure~\ref{fig:sim_confusion} illustrates the confusion matrix
from the first replication (ARI $= 0.977$),
in which cluster assignment is near-perfect:
Cluster~$k=3$ (low-ability) is perfectly recovered
($100\%$ correct), while Clusters~$k=1$ and $k=2$ are each
assigned correctly for 98\% of students,
with only one misclassified observation per cluster.

\subsubsection{Key Parameter Recovery}
\label{subsubsec:sim_theta}

Table~\ref{tab:sim_proposed_theta} summarizes ability ($\theta$), drift ($\delta$), and difficulty ($b$) recovery across the 100 replications.

\paragraph{Ability trajectory recovery.}
For ability trajectories, the proposed model achieved a mean $\mathrm{MAB}(\theta) = 0.487$ logit units ($\mathrm{SE} = 0.003$) and a mean $\mathrm{MSE}(\theta) = 0.452$ ($\mathrm{SE} = 0.005$).
The 95\% HPD coverage probability was $0.953$ ($\mathrm{SE} = 0.001$), closely matching the nominal 0.95 level.

\begin{table}
\footnotesize
  \centering
  \caption{Parameter Recovery of the Proposed Model Across 100 Replications}
  \label{tab:sim_proposed_theta}
  \begin{threeparttable}
    \begin{tabular}{lcccc}
      \toprule
      Metric & Mean & Median & SD & SE \\
      \midrule
      $\mathrm{MAB}(\theta)$ & 0.487 & 0.483 & 0.025 & 0.003 \\
      $\mathrm{MSE}(\theta)$ & 0.452 & 0.444 & 0.051 & 0.005 \\
      $\mathrm{CP}_{95\%}(\theta)$ & 0.953 & 0.954 & 0.008 & 0.001 \\
      $\mathrm{MAB}(\delta)$ & 0.102 & 0.097 & 0.029 & 0.003 \\
      $\mathrm{MSE}(\delta)$ & 0.025 & 0.020 & 0.014 & 0.001 \\
      $r(\delta)$& 0.933 & 0.940 & 0.033 & 0.003 \\
      $\mathrm{MAB}(b)$& 0.194 & 0.194 & 0.013 & 0.001 \\
      $\mathrm{CP}_{95\%}(b)$ & 0.921 & 0.922 & 0.022 & 0.002 \\
      \bottomrule
    \end{tabular}
  \end{threeparttable}
\end{table}

\begin{figure}
\centering
\includegraphics[scale=0.6]{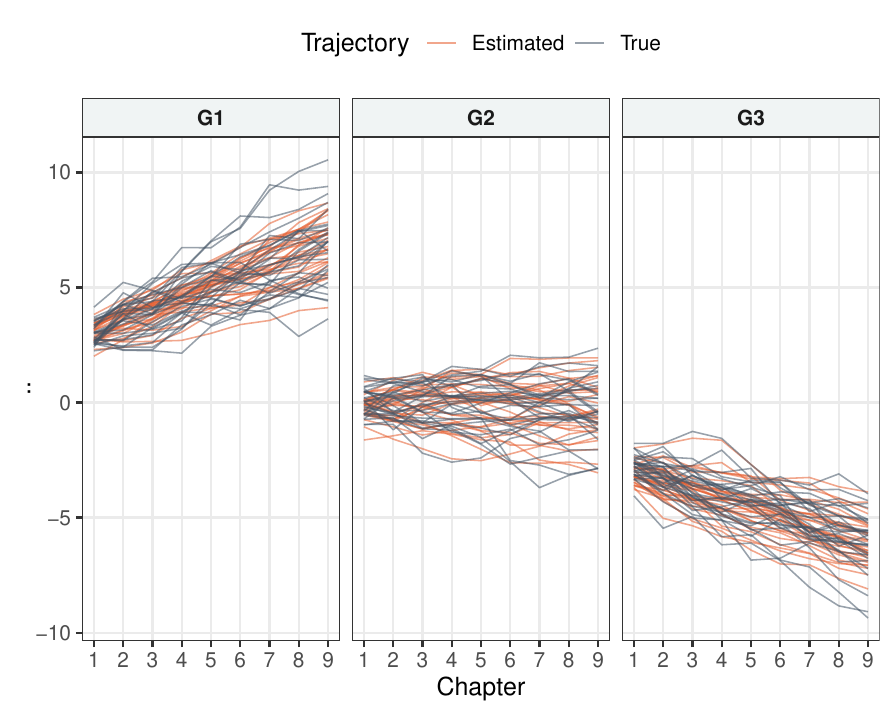}
\caption{\label{fig:sim_theta_traj}Individual Ability Trajectories: Estimated versus True Values by Cluster in a Representative Replication.}
\end{figure}

The quantitative recovery indices are complemented by
Figure~\ref{fig:sim_theta_traj}, which displays the estimated
(red) and true (blue) $\theta$ trajectories for 30 randomly
selected students per cluster across $t = 1, \ldots, 9$ in
a representative replication.
Three cluster-specific patterns emerge.

(1) Cluster $G1$ (high-ability; positive drift).
Initial abilities are concentrated in $\theta \in [2,\,4]$
and increase monotonically to $\theta \in [5,\,10]$ by
chapter~9, consistent with the positive drift generated by
$\beta_{1,1}^* = \beta_{2,1}^* = +0.15$.
Estimated trajectories track the true trajectories closely
throughout the observation window, with only minor
chapter-level deviations, reflecting the model's high
precision for students whose ability signal is both strong
and directionally consistent.

(2) Cluster $G2$ (average-ability; zero drift).
True trajectories fluctuate around $\theta \approx 0$
within an approximate range of $[-4,\,+2]$, with no
systematic upward or downward trend, as dictated by the
zero drift coefficients ($\beta_{1,2}^* = \beta_{2,2}^* = 0$).
Estimated and true trajectories remain closely aligned
across most time points, though mild divergences emerge
for some individuals at later chapters ($t \geq 6$),
plausibly attributable to between-chapter item pool
heterogeneity accumulating over the course.

(3) Cluster $G3$ (low-ability; negative drift).
Starting from $\theta \in [-2,\,-3]$, true trajectories
decline markedly, reaching approximately $-10$ for several
students by chapter~9, a pattern consistent with the
negative drift coefficients
($\beta_{1,3}^* = \beta_{2,3}^* = -0.15$).
The between-student dispersion in true trajectories widens
substantially over time (larger inter-individual spread at
$t = 9$ than at $t = 1$), reflecting cumulative process
noise compounded over eight transitions.
Estimated trajectories capture the overall downward trend
faithfully; however, localized over- or under-estimation
is somewhat more pronounced in $G3$ than in $G1$,
a consequence of the larger absolute $\theta$ range
and the reduced information content of binary items
at extreme ability levels \citep{embretson2013item}.

Overall, the estimated trajectories reproduce the
directional and magnitude characteristics of the true
ability dynamics across all three clusters, with $G1$
showing the sharpest recovery and $G3$ displaying the
greatest residual variability at later time points.
This pattern is consistent with the theoretical expectation
that posterior precision diminishes as ability departs
further from the item pool's mean difficulty
\citep{Lord1980}.

\begin{figure}
\centering
\includegraphics[scale=0.5]{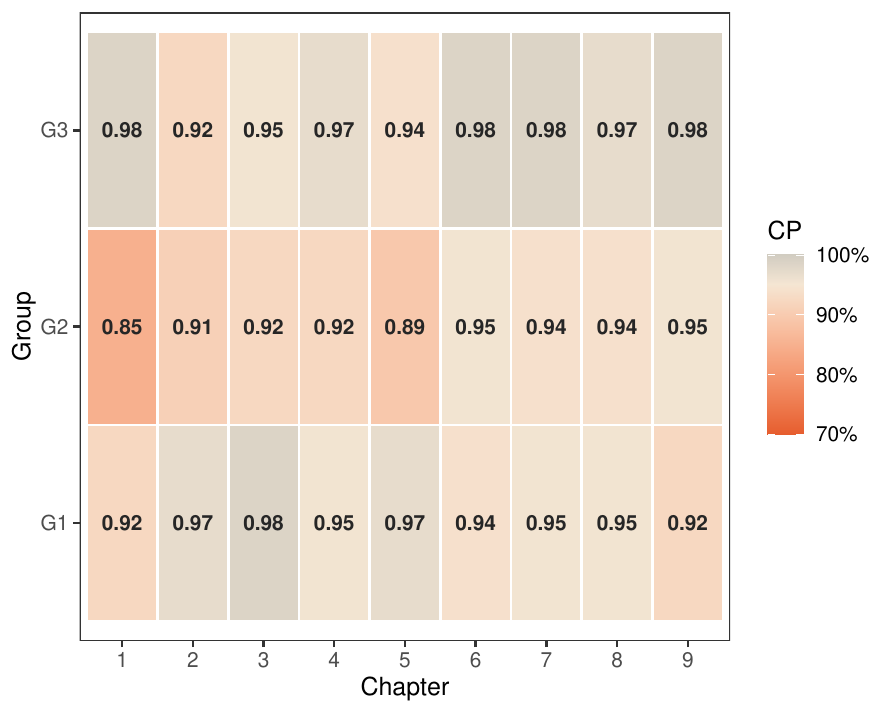}
\caption{\label{fig:sim_cp}Group $\times$ Chapter
    Heatmap of 95\% HPD Coverage Probabilities for
    $\theta_{it}$ in a Representative Replication.}
\end{figure}

The near-nominal coverage is further illustrated in
Figure~\ref{fig:sim_cp}, which displays the
group $\times$ chapter CP heatmap from the first replication.
Coverage probabilities are distributed in $[0.92,\,0.98]$
for clusters~$G1$ and $G3$, with a single isolated low value
at $G2$, $t=1$ ($\mathrm{CP} = 0.85$).
This isolated deviation reflects higher posterior uncertainty
for the zero-drift cluster at the initial time point before
engagement data accumulate, and does not indicate systematic
under-coverage.
The overall mean CP in the representative replication is
$0.947$, consistent with the 100-replication aggregate of
$0.953$.

\paragraph{Drift recovery.} For drift recovery, since the cluster-specific scaling coefficients
$\beta_{p,k}$ and the global spline basis coefficients
$\mathbf{b}_p$ are scale-unidentifiable in isolation, the \textit{composite drift} $\delta_{it}$ is adopted as the primary identifiable target for evaluation.
The overall drift correlation of $r(\delta) = 0.933$
confirms that the B-splines specification successfully
captures the underlying $x^3$ and $\exp(x)$ functional
form from observed binary responses.

\paragraph{Item difficulty recovery.} Item difficulty recovery was satisfactory:
$\mathrm{MAB}(b) = 0.194$ and $\mathrm{CP}_{95\%}(b) = 0.921$, a slight shortfall from the nominal 0.95 consistent with the known tendency of HPD intervals to be marginally liberal in high-dimensional IRT models \citep{fox2010bayesian}.

\section{Empirical Application}

\label{sec:empirical}

We apply the proposed model to longitudinal item response
data from the CourseKata online learning platform \citep{CourseKata2023}. CourseKata is an interactive online textbook for statistics and data science that students access through their institution's learning management system, such as Canvas or Google Classroom. The platform brings together three complementary forms of data. First, it collects self-report survey data, including chapter-level pulse checks that capture students' motivational beliefs and how these beliefs fluctuate during learning. Second, it records behavioral log data that document students' engagement with course materials as they move through each chapter. Third, it captures performance data from more than 1,500 formative assessment items embedded throughout the instructional materials across 12 chapters.

Because these measures are integrated directly into the CourseKata learning environment and collected across a wide range of classroom and institutional contexts---including high schools, community colleges, and two- and four-year colleges---CourseKata provides a distinctive research infrastructure for studying theoretically motivated learning processes in authentic educational settings \citep{zhang2024representational}.

The analysis pursues three substantive goals:
(i)~to determine how many distinct learner profiles are
supported by the data without a prespecified $K$,
(ii)~to characterize each profile in terms of ability level
and trajectory across chapters, and
(iii)~to evaluate whether engagement quantity predicts
within-individual ability drift.

\subsection{Data and Measures}
\label{subsec:data}

Data were collected during the Winter 2023 semester from
undergraduate students enrolled in an introductory statistics and data science course at a North American university.
The course was delivered entirely online via the CourseKata
platform, a research-grade learning management system that
records fine-grained student interaction logs alongside
item-level responses \citep{CourseKata2023}.
Of 213 students who began the course, 198 (93.0\%) with complete background information were retained for analysis.

The course comprised $T = 9$ instructional chapters.
Chapter-level item counts ranged from 12 to 32
($J_t \in \{13,\,32,\,31,\,32,\,32,\,30,\,12,\,20,\,29\}$),
for a total of $\sum_t J_t = 230$ dichotomous scored items.

Two chapter-level engagement quantity measures were
constructed per student from the platform interaction logs:
\begin{enumerate}
  \item \textit{Average session duration}($x_{it}^{(1)}$, \texttt{dmean}):
    the mean duration (in minutes) per active reading
    session in chapter $t$, standardized to have mean zero
    and unit variance across all ($i$, $t$) pairs.
  \item \textit{Session count}
    ($x_{it}^{(2)}$, \texttt{ns}):
    the number of distinct reading sessions in chapter $t$,
    standardized similarly.
\end{enumerate}

\paragraph{Background variables.}
Four background variables were measured via a pre-course survey
and used post-hoc to characterise cluster profiles.
Each variable was originally recorded on a multi-category ordinal
or nominal scale and subsequently dichotomised into a binary
indicator (0/1) for 
the reason that these variables serve solely as \textit{post-hoc
descriptors} rather than model inputs, a binary summary
provides a readily interpretable and theoretically meaningful
threshold for cluster comparison without imposing unwarranted
assumptions about equal spacing between ordinal categories
\citep{Jamieson2004}.

The four variables, together with their original response scales
and dichotomisation rules, are defined as follows:

\begin{enumerate}

  \item \textit{Math anxiety.} Measured on a six-point Likert scale ranging from
    \textit{Strongly disagree} (1) to \textit{Strongly agree} (6).
    Categories 4--6 were coded 1 (endorses math anxiety);
    all remaining responses were coded 0.

  \item \textit{Memorization belief.}
    Measured on the same six-point Likert scale.
    Agreement that statistics is primarily learned through
    memorisation (categories 4--6) was coded 1; disagreement (categories 1--3) was
    coded 0.

  \item \textit{Pre-course mathematics performance.}
    Measured on a five-point scale:
    \textit{Not at all well} (1), \textit{Slightly well} (2),
    \textit{Moderately well} (3), \textit{Very well} (4),
    \textit{Extremely well} (5).
    Students who reported Categories 4--5 were classified
    as high prior mathematics performers and coded 1; all others were coded 0.

  \item \textit{Coding experience.}
    Measured on a four-category nominal scale:
    \textit{No} (1),
    \textit{Self-taught, no formal course} (2),
    \textit{Completed a programming course} (3),
    \textit{Used programming in a non-programming course} (4).
    Any prior exposure to programming---formal or
    informal---was coded 1
    (categories 2--4); no experience was coded 0.

\end{enumerate}

Together with gender (female coded 1), all these five binary indicators were excluded from the dynamic IRT
model itself and reserved solely for post-hoc characterisation
of the identified clusters, thereby avoiding any circularity
between the clustering procedure and the background-variable
profiles used to interpret it.



\subsection{Results}
\label{subsec:results}

\subsubsection{Cluster Enumeration and Learner Profiles. }
\label{subsubsec:k_result}

\begin{figure}
\centering
\includegraphics[scale=0.5]{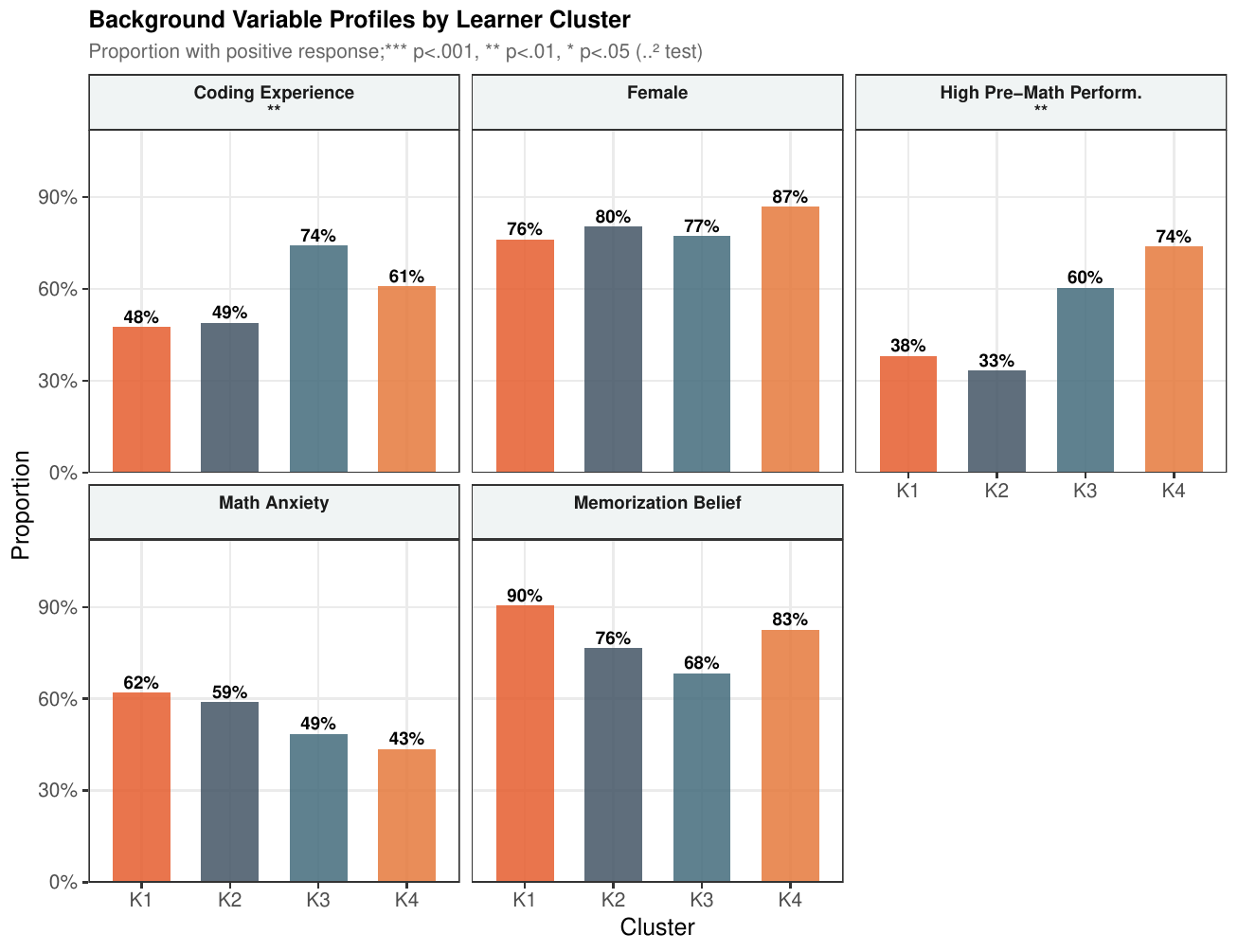}
\caption{\label{fig:background} Background Variable Profiles by Learner Cluster.}
\end{figure}

The estimated number of clusters is $\hat{K}=4$ with the proposed model, indicating that the students can be categorized into 4 groups according to different types of ability trajectories. 
Table~\ref{tab:cluster_summary} reports the posterior summaries of cluster-specific parameters.
Clusters are labeled K1-K4 in ascending order of mean
ability. Background variable profiles among these four groups are shown in Figure~\ref{fig:background}. 

\begin{table}
\footnotesize
  \centering
  \caption{Cluster Summary: Posterior Means and 95\% HPD Intervals}
  \label{tab:cluster_summary}
  \renewcommand{\arraystretch}{1.30}
  \begin{threeparttable}
    \begin{tabular}{lrrrrrr}
      \toprule
      Cluster& $n$ & \%
        & $\hat{\theta}_{m,k}$ [95\% HPD]
        & $\hat{\theta}_{i,1}$\tnote{a}
        & $\hat{\theta}_{i,9}$\tnote{a}
        & $\widehat{\Delta\theta}$\tnote{b} \\
      \midrule
      K1 (Struggling--Declining)
        & 21 & 10.6
        & 1.07 [0.83,\;1.33]
        & 1.16 & 1.04 & $-$0.12\\
      K2 (Low--Stable)
        & 51 & 25.8
        & 1.91 [1.69,\;2.11]
        & 1.97 & 1.89 & $-$0.08 \\
      K3 (Mainstream--Stable)
        & 103 & 52.0
        & 2.80 [2.64,\;3.01]
        & 2.77 & 2.83 & $+$0.06 \\
      K4 (High--Improving)
        & 23 & 11.6
        & 4.21 [3.71,\;4.72]
        & 4.10 & 4.29 & $+$0.19 \\
      \bottomrule
    \end{tabular}
    \begin{tablenotes}
      \scriptsize
      \item \textit{Note. }95\% HPD = highest posterior density interval.\\
      All cluster mean ability values are on the logit
        (log-odds) scale.\\
      \item[a] Posterior mean of the cluster-average ability
        at Chapter 1 and Chapter 9, respectively,
        computed as the mean of $\hat{\theta}_{it}$ across
        students assigned to the cluster.\\
      \item[b] $\widehat{\Delta\theta} =
        \bar{\hat\theta}_{k,9} - \bar{\hat\theta}_{k,1}$;the cluster-average ability change from Chapter~1
        to Chapter~9.
    \end{tablenotes}
  \end{threeparttable}
\end{table}

According to Table~\ref{tab:cluster_summary} and Figure~\ref{fig:background}, four substantively distinct profiles emerged:

\paragraph{Cluster K1: Struggling--Declining (10.6\%).}
The smallest cluster, comprising 21 students, exhibited
the lowest mean ability ($\hat{\theta}_{m,1} = 1.07$,
95\% HPD: [0.83, 1.33]) and a modest negative trajectory
($\widehat{\Delta\theta} = -0.12$).
Students in this group appeared to fall progressively
further behind as course content became more demanding.
Their high rate of math anxiety (62\%) and low prior mathematics performance (38\% high pre-course
performance) and low coding experience (48\%) suggest that inadequate mathematical
preparation and coding experience may have compounded their difficulties
with accumulating statistical concepts.

\paragraph{Cluster K2: Low--Stable (25.8\%).}
51 students formed a low-ability but
largely stable group ($\hat{\theta}_{m,2} = 1.91$,
95\% HPD: [1.69, 2.11]; $\widehat{\Delta\theta} = -0.08$).
The near-zero trajectory indicates that, while these
students consistently answered items at a below-average
level, their relative standing did not deteriorate
further over the course.

\paragraph{Cluster K3: Mainstream--Stable (52.0\%).}
The modal cluster of 103 students---representing more
than half the sample---exhibited average ability
($\hat{\theta}_{m,3} = 2.80$,
95\% HPD: [2.64, 3.01]) and an essentially flat trajectory
($\widehat{\Delta\theta} = +0.06$).
This group closely resembles the ``adequate but
unexceptional'' learner archetype identified in
introductory statistics research
\citep{Garfield2008}.

\paragraph{Cluster K4: High--Improving (11.6\%).}
23 students demonstrated both the highest mean
ability ($\hat{\theta}_{m,4} = 4.21$,
95\% HPD: [3.71, 4.72]) and the only clearly positive
trajectory across the course
($\widehat{\Delta\theta} = +0.19$).
Their high prior mathematics performance rate (74\%)
suggests that these students entered with strong
quantitative foundations that continued to compound
as course content progressed.

\begin{figure}
\centering

\begin{minipage}{0.90\textwidth}
\centering
\includegraphics[width=0.7\linewidth]{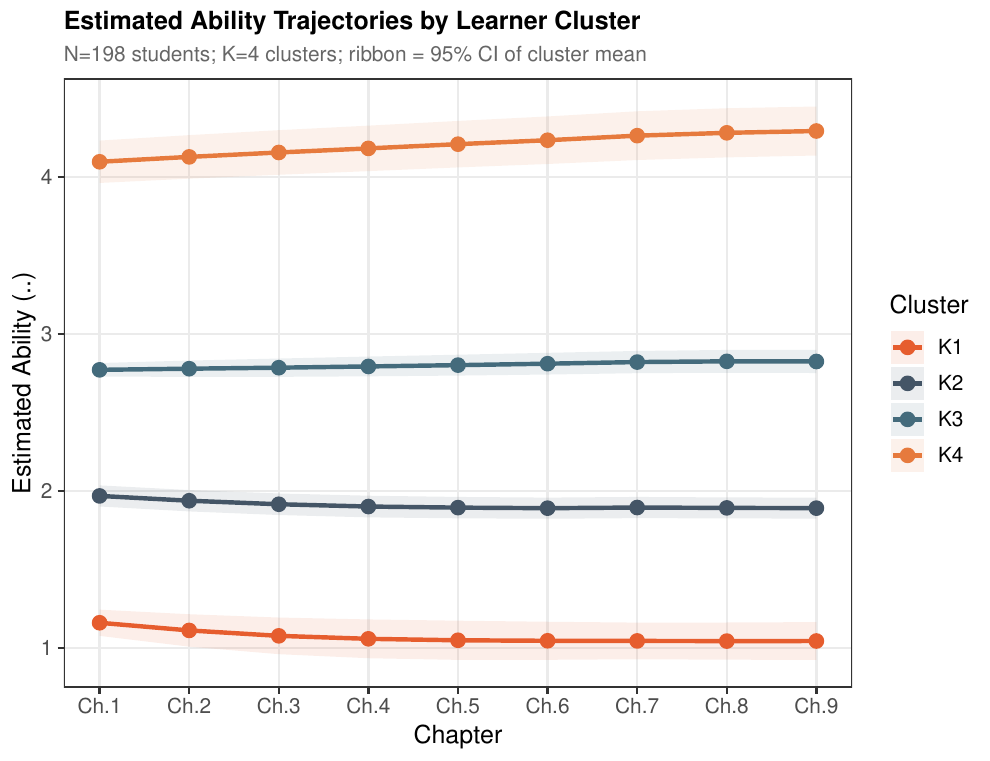}

\vspace{1mm}
\textbf{A.} Posterior mean ability trajectories by learner cluster
across nine chapters.
\end{minipage}

\vspace{4mm}

\begin{minipage}{0.90\textwidth}
\centering
\includegraphics[width=0.7\linewidth]{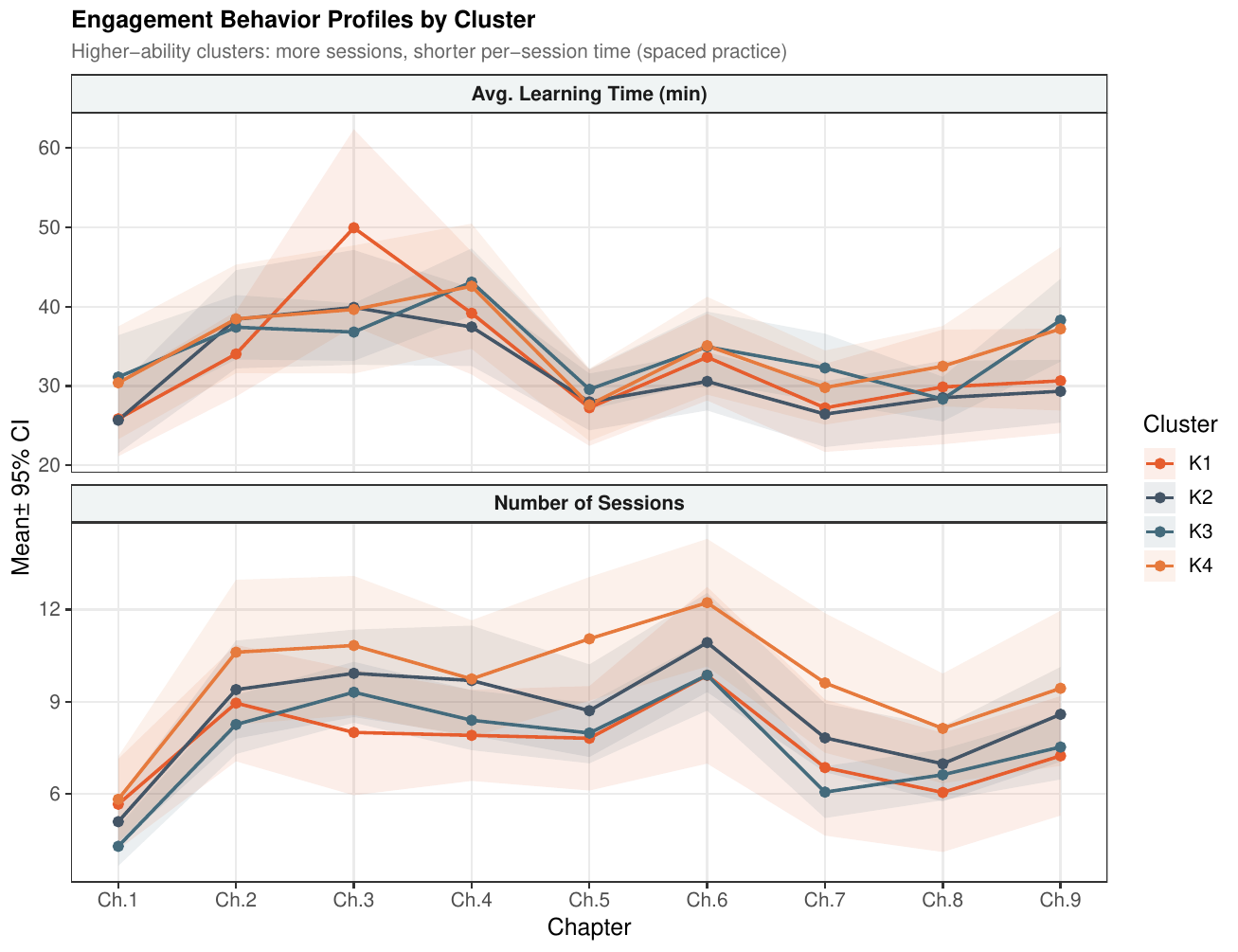}

\vspace{1mm}
\textbf{B.} Posterior estimates of nonlinear engagement-drift functions
by cluster.
\end{minipage}

\caption{\label{fig:trajectory_engagement}
Posterior summaries by learner cluster. Panel A shows posterior mean
ability trajectories across the nine chapters. Panel B shows posterior
estimates of nonlinear engagement-drift functions by cluster.}
\end{figure}

\paragraph{Overall ability stability.}
A striking feature of all four profiles is the
high degree of longitudinal stability.
The estimated process precision was
$\hat{\tau}_\theta = 36.33$
(95\% HPD: [28.14, 44.27]),
corresponding to a process standard deviation of
$\hat{\sigma}_\theta = 1/\sqrt{\hat{\tau}_\theta}
\approx 0.166$ logit units.
To contextualize this magnitude: the four cluster means
span approximately 3.1 logit units, meaning that
chapter-to-chapter stochastic variation within an
individual represents only about 5\% of the total
between-cluster ability range (see Figure \ref{fig:trajectory_engagement}(A)).

\subsubsection{Engagement Covariates and Ability Drift}
\label{subsubsec:engagement}

Table~\ref{tab:hyperparams} reports posterior summaries
for the global spline coefficients
$\mathbf{b}_1$(\texttt{dmean}) and
$\mathbf{b}_2$ (\texttt{ns}),
and the cluster-specific scaling coefficients
$\{\beta_{p,k}\}$.

\begin{table}
\footnotesize
  \centering
  \caption{Posterior Summaries: Hyperparameters and Cluster-Level Engagement Coefficients}
  \label{tab:hyperparams}
  \renewcommand{\arraystretch}{1.25}
  \begin{threeparttable}
  \begin{tabular}{llrrrr}
    \toprule
    Type & Parameter& Mean & SD& HPD\textsubscript{lo} & HPD\textsubscript{hi}\\
    \midrule
    \multicolumn{6}{l}{\textit{Panel A: Global spline coefficients}} \\[3pt]
    Spline (\texttt{dmean})
      & $b_{11}$ & 0.003 & 0.286 & $-$0.578 & 0.566 \\
    Spline (\texttt{dmean})
      & $b_{12}$ & $-$0.010 & 0.342 & $-$0.693 & 0.665 \\
    Spline (\texttt{dmean})
      & $b_{13}$ & $-$0.049 & 0.460 & $-$0.939& 0.897 \\
    Spline (\texttt{ns})
      & $b_{21}$ & $-$0.012 & 0.257 & $-$0.503 & 0.476 \\
    Spline (\texttt{ns})
      & $b_{22}$ & 0.024 & 0.287 & $-$0.560 & 0.608 \\
    Spline (\texttt{ns})
      & $b_{23}$ & $-$0.018 & 0.248 & $-$0.531 & 0.500 \\
    \midrule
    \multicolumn{6}{l}{\textit{Panel B:
      Cluster-specific scaling coefficients}} \\[3pt]
    $\beta_1$ (\texttt{dmean})
      & K1 & $-$0.034 & 1.514 & $-$2.929 & 3.286 \\
    $\beta_1$ (\texttt{dmean})
      & K2 & 0.032 & 1.249 & $-$2.487 & 2.670 \\
    $\beta_1$ (\texttt{dmean})
      & K3 & 0.053 & 0.966 & $-$1.957 & 2.012 \\
    $\beta_1$ (\texttt{dmean})
      & K4 & $-$0.024 & 1.724 & $-$3.300 & 3.340 \\
    $\beta_2$ (\texttt{ns})
      & K1 & 0.010 & 1.544 & $-$3.248 & 2.971 \\
    $\beta_2$ (\texttt{ns})
      & K2 & $-$0.030 & 1.112 & $-$2.487 & 2.670 \\
    $\beta_2$ (\texttt{ns})
      & K3 & 0.072 & 1.145 & $-$2.123 & 2.539 \\
    $\beta_2$ (\texttt{ns})
      & K4 & $-$0.045 & 1.774 & $-$3.710 & 3.295 \\
    \bottomrule
  \end{tabular}
  \begin{tablenotes}
    \scriptsize
    \item \textit{Note.}
      All engagement covariates were standardized prior to
      B-spline basis construction.
      Spline coefficients $b_{sl}$ are on the
      normalized basis scale.\end{tablenotes}
  \end{threeparttable}
\end{table}

All global spline coefficients and cluster-specific
scaling factors had posterior means close to zero
with 95\% HPD intervals that comfortably spanned zero
(Table~\ref{tab:hyperparams}, Panel~A--B).
The estimated engagement-drift functions
$\hat{f}_1(\cdot)$ and $\hat{f}_2(\cdot)$ were
therefore essentially flat across the full covariate
support---a pattern that held uniformly across all
four learner clusters (Figure~\ref{fig:trajectory_engagement}(B)).

\subsubsection{Item Parameters}
\label{subsubsec:item}

Table~\ref{tab:item_params} reports chapter-level
summaries of the item difficulty parameters.

\begin{table}
\footnotesize
  \centering
  \caption{Item Difficulty Parameter Summaries by Chapter}
  \label{tab:item_params}
  \renewcommand{\arraystretch}{1.25}
  \begin{threeparttable}
    \begin{tabular}{lrrrrrr}
      \toprule
      Chapter
        & $J_t$
        & $\hat{\bar{b}}_t^{\mathrm{raw}}$
        & $\hat{\sigma}_{b,t}$
        & $\hat{b}_{t,\min}$
        & $\hat{b}_{t,\max}$\\
      \midrule
      Ch.1 & 13& 0.028 & 1.64 & $-$1.61 & 4.40\\
      Ch.2 & 32 & 0.013 & 1.90 & $-$4.21 & 4.24\\
      Ch.3 & 31 & $-$0.031 & 1.95 & $-$4.51 & 4.20\\
      Ch.4 & 32 & $-$0.023 & 1.87 & $-$4.12 & 4.22\\
      Ch.5 & 32 & 0.028 & 1.87 & $-$4.06 & 4.28\\
      Ch.6 & 30 & 0.002 & 1.99 & $-$4.50 & 4.27\\
      Ch.7 & 12 & $-$0.013 & 1.72 & $-$1.74 & 4.42\\
      Ch.8 & 20 & 0.017 & 1.77 & $-$2.68 & 4.37\\
      Ch.9 & 29 & 0.020 & 1.75 & $-$3.51 & 4.33\\
      \midrule
      Overall & 230 & 0.004 & 1.83 & $-$4.51 & 4.42\\
      \bottomrule
    \end{tabular}
    \begin{tablenotes}
      \scriptsize
      \item \textit{Note.}
        $J_t$ = number of items in chapter $t$.
        $\hat{\bar{b}}_t^{\mathrm{raw}}$ = posterior
        mean of the uncentered chapter mean difficulty,
        interpreted as absolute chapter-level difficulty
        relative to the overall mean.
        $\hat{\sigma}_{b,t}$ = posterior mean of the
        chapter-specific item difficulty spread.
        $\hat{b}_{t,\min}$ and $\hat{b}_{t,\max}$ =
        minimum and maximum posterior mean item
        difficulties (centered within chapter).
    \end{tablenotes}
  \end{threeparttable}
\end{table}

Chapter-level absolute difficulty values
$\hat{\bar{b}}_t^{\mathrm{raw}}$ were uniformly close
to zero (range: $-$0.031 to $+$0.028), indicating that
the nine chapters were well-calibrated in terms of
overall difficulty and that the strong longitudinal
ability stability observed in learner profiles cannot be attributed to compensatory difficulty changes across chapters.
The item difficulty spread $\hat{\sigma}_{b,t}$
was consistent across chapters
($\hat{\sigma}_{b,t} \in [1.64, 1.99]$), and all
chapters contained both very easy and very difficult items, confirming adequate item difficulty coverage throughout the
course.

\section{Discussion}

\label{sec:discussion}

This study introduced a hierarchical Bayesian framework that simultaneously identifies the number of
latent learner clusters, estimates nonparametric engagement-ability drift functions, and tracks individual ability trajectories across instructional chapters. 
We evaluated the proposed approach using simulation
studies and longitudinal item-response data from 198 undergraduates
enrolled in an online introductory statistics course.
The analysis yielded three principal findings: the emergence of four substantively meaningful learner profiles, the striking stability of ability trajectories across chapters, and the absence of measurable engagement-drift relationships.
We then reflect on the methodological contributions of the proposed framework, and outline directions for future research.

\subsection{Learner Profiles and the Case for Four Clusters}
\label{subsec:discuss_profiles}

Our approach identified $\hat{K} = 4$ latent learner profiles without requiring pre-specification of $K$.
The four profiles: Struggling-Declining, Low-Stable, Mainstream-Stable, and
High-Improving, closely correspond to archetypes
documented in the introductory statistics education
literature \citep{Garfield2008, Zieffler2015}.
The predominance of the Mainstream-Stable group
is consistent with prior evidence that many
introductory statistics students occupy a middle range of performance \citep{Nasser2004}, while
the small Struggling-Declining group reflects a subgroup for whom introductory statistics
may represent a substantial academic barrier \citep{onwuegbuzie2004academic}.

\subsection{Ability Stability: Crystallized Competency in
  Online Statistics Education}
\label{subsec:discuss_stability}

The most striking finding of the empirical analysis was
the extremely high longitudinal ability stability.
The estimated process precision $\hat{\tau}_\theta$
implies that chapter-to-chapter stochastic variation in individual ability represents only about 5\% of the total between-cluster ability range.
Thus, students' relative positions were largely
preserved across the nine chapters.

This result is consistent with the theoretical framework
of crystallized intelligence \citep{Cattell1971},
which distinguishes between fluid abilities and crystallized abilities that reflect accumulated knowledge and are less
sensitive to short-term change.  
Introductory statistics achievement, particularly as
measured by embedded reading comprehension items of the
type used in CourseKata, may therefore reflect relatively stable mathematical
and quantitative preparation more than short-term changes in engagement
during a single semester.
The implication for educational practice is that interventions focused solely on increasing
in-course engagement quantity, such as time-on-task or session
frequency, may be insufficient for students who enter with weak
mathematical preparation.
More promising avenues may include prerequisite
curriculum design and course placement
procedures that identify and support at-risk learners
\emph{before} enrollment rather than attempting to
remediate during instruction.

\subsection{Why Engagement Quantity Did Not Predict Drift}
\label{subsec:discuss_engagement}

All the estimated
B-splines functions $\hat{f}_1(\cdot)$ and
$\hat{f}_2(\cdot)$ were flat across the full covariate
support.
Three non-mutually-exclusive explanations merit consideration.

First, and most structurally, the high process precision
$\hat{\tau}_\theta$ leaves very little statistical room
for any covariate to explain drift.
When the true ability trajectories are nearly constant,
the data contain limited information about
within-student change, and the posterior for engagement effects is
therefore pulled toward zero.
This is not a failure of the B-splines specification;
rather, it reflects the limited evidence for
within-individual ability drift in these data.

Second, the engagement measures available in this study
were exclusively \emph{quantity}-based (time, session
count).
A substantial body of evidence in cognitive psychology
suggests that learning outcomes are more strongly
predicted by the \emph{quality} of engagement---in
particular, retrieval practice, interleaving, and
elaborative interrogation---than by the raw amount of
time spent \citep{Dunlosky2013, Karpicke2012}.
Platform-level engagement logs of the type analyzed here do not capture whether students engaged in active
self-testing, re-read strategically, or sought out
supplementary resources; the absence of engagement
effects on drift may therefore reflect measurement
inadequacy rather than a true null relationship.

Third, the course design itself may have introduced
range restriction in engagement: CourseKata requires
students to complete all items before advancing, creating
a floor on engagement quantity that compresses
between-student variability and attenuates any
underlying engagement-achievement association \citep{musaji2020long}.




\subsection{Methodological Contributions}
\label{subsec:discuss_methods}

Beyond the substantive findings, the proposed
framework makes two methodological contributions to the
psychometric literature.

First, the integration of the MFM prior into a dynamic
IRT model demonstrates that automatic cluster enumeration
is feasible without sacrificing the state-space structure
that enables ability trajectory estimation.
This extends the use of Bayesian nonparametric methods
in educational measurement to longitudinal IRT settings
\citep{miller2018mixture,wang2020longitudinal,castro2024time}.

Second, the B-splines drift specification provides a
flexible diagnostic tool for applied researchers: when
engagement effects are present in the data, the
estimated functions $\hat{f}_p(\cdot)$ reveal their
shape without parametric constraints; when effects are
absent, the functions collapse to zero without
generating spurious nonlinear artifacts, as confirmed
by the simulation study. 
This behavior is useful in applied settings where the
presence and shape of engagement effects are unknown.


\subsection{Future Directions}
\label{subsec:Future}

There are several directions for future research.

First, \textit{engagement quality metrics} should be incorporated as predictors
of ability drift.  Retrieval practice frequency, interleaving
indicators, spacing patterns, response revision behavior, and self-testing
behavior are all quantifiable from platform logs and are theoretically motivated
by the science of learning \citep{Dunlosky2013}. Including these variables would
help distinguish productive engagement from more superficial forms of activity,
such as simply spending more time on the platform. 

Second, \textit{cross-institutional replication} using the multi-site CourseKata
dataset \citep{CourseKata2023} would establish whether the four-profile solution is
stable across universities differing in selectivity, student demographic
composition, instructional modality, and course design. 
Such work would clarify whether the identified profiles
reflect general learner patterns or local instructional and assessment
contexts.
This would strengthen evidence for the generalizability of the proposed
framework and support its use in broader educational contexts.

Third, \textit{experimental designs} in which a targeted learning intervention is
randomized at the student level would provide cleaner causal leverage on the
engagement--ability relationship within the proposed framework. For example,
students could be randomly assigned to receive prompts for retrieval practice,
adaptive review schedules, or structured self-testing activities, and these intervention indicators could be incorporated as
predictors of ability drift or latent profile transitions.

Fourth, \textit{model extensions} accommodating multidimensional ability
structures \citep{reckase2009multidimensional}, polytomous responses, and time-varying item
parameters \citep{castro2024time} would broaden the applicability of the framework to a
wider class of online learning contexts and assessment designs. 
These extensions would allow the framework to handle
multiple competencies, partial-credit outcomes, and changes in item
functioning across contexts.

\bibliographystyle{plainnat}
\bibliography{example}

\end{document}